\documentclass[12pt]{iopart}
\usepackage{graphicx}
\usepackage{amssymb}

\begin{document}
	
	\title[Grav. Waves beyond the Linear Approximation and Grav. Wave Reflection]{Gravitational Waves beyond the Linear Approximation and Gravitational Wave Reflection}
	
	\author{Victor Atanasov}	
	\address{Department of Condensed Matter Physics and Microelectronics, Faculty of Physics, Sofia University "St. Kliment Ohridski", 5 blvd J. Bourchier, Sofia 1164, Bulgaria}
	\ead{vatanaso@phys.uni-sofia.bg}
	
	\author{Avadh Saxena}
	\address{Theoretical Division and Center for Nonlinear Studies, Los Alamos National Laboratory, Los Alamos, NM 87545 USA}
	\ead{avadh@lanl.gov}

	\begin{abstract}
		We derive a relativistic field equation for the trace of the metric perturbation beyond the weak field approximation to the Einstein field equations. The dynamics is governed by a massive Klein-Gordon equation on curved space-time, where the effective mass of the field is associated with the material and the dark energy content via the cosmological term. We solve the equation in the case of a Schwarzschild black hole and show that it can be cast into an effective Schr\"odinger form with an effective geometric potential which binds the zero angular momentum states. The non-zero angular momentum states experience a positive potential peak before the event horizon pointing to gravitational waves scattering. Black holes scatter gravitational waves and thus we provide an unambiguous testable prediction of black hole existence. The Newtonian limit for this equation points to the possibility of reflecting gravitational waves at interfaces with sharp density boundary, thus opening up gravitational wave propulsion physics. We discuss this type of propulsion in the light of Newton's third law of Mechanics. Compelling questions such as the existence of quanta of this field which may account for the dark matter content are also addressed.
	\end{abstract}

\section{Introduction}

General Theory of Relativity encompasses the Equivalence Principle \cite{AE1} (or Einstein’s happiest thought \cite{AE2}) and the Einstein field equations \cite{EFE}. It is by far the best theory of gravitation we have \cite{GPB, GW1}.

Einstein field equations are built on the pillars of mathematical consistency and the supposition that some physical quantity should be conserved. As usual, it is the energy and momentum \cite{MTW}. In this case, it is the geometrical field represented by the Einstein tensor $G_{\sigma \nu}$ that is related to the stress-energy tensor $T_{\sigma \nu}$ for the sole reason that both have vanishing divergence $G_{\sigma \nu} = \kappa T_{\sigma \nu}$ \cite{SC}. The Einstein gravitational coupling constant $\kappa$ balances the dimensions of the physical quantities on the sides of the equation and is very small $\kappa ={8\pi G}/{c^{4}}\approx 2\times 10^{-43}$ N$^{-1}$. With this choice of units, the stress-energy tensor's components are given in terms of energy density.

For the purposes of the present exploration the following form of the Einstein field equations is more appropriate and revealing
\begin{eqnarray}\label{EinsteinFE}
	R_{\sigma \nu} = \kappa \left( T_{\sigma \nu} - \frac12 g_{\sigma \nu} T\right) + g_{\sigma \nu} \Lambda.   
\end{eqnarray}    
The trace of the stress-energy tensor is given by $T=T^{\alpha}_{\; \alpha}$. Here the Ricci curvature tensor is denoted by $R_{\sigma \nu}$, the metric tensor by $g_{\sigma \nu}$ and the cosmological dark energy density term by $\Lambda$. Hereinafter, we refer to the right-hand side of equation (\ref{EinsteinFE}) as the  ``material content tensor'' 
\begin{eqnarray}\label{MC_tensor}
	M_{\sigma \nu} = \kappa \left( T_{\sigma \nu} - \frac12 g_{\sigma \nu} T\right) + g_{\sigma \nu} \Lambda.   
\end{eqnarray}    

The paper deals with an extension of the popular weak-field limit of the Einstein field equations. The linearized gravitation is centered around the flat space-time Minkowski metric and explores the consequences for a small perturbation to that metric in the presence of matter. This construction is very limited and requires a gauge fixing \cite{LG}. There are also issues which arise from the very form of the equations of motion (Lorentz gauge) $ \square  h_{\mu \nu} \propto T_{\mu \nu }$, that is a quantity in the 1st order in the small parameter ($\square  h_{\mu \nu}$) equals a quantity in the 0th order (the stress-energy tensor). Naturally, the 0th order quantity on the right-hand side of the equations should lead to the solution for the ``background'' metric, which in the case of linearized gravitation is the Minkowski metric, and as a result certain left and right hand side terms including $T_{\mu \nu }$ should cancel out.


Besides the linearized gravitation centered around the Minkovski metric, there is an alternative approach to gravitational radiation which dates back to the 1960's and is referred to as the Newman-Penrose formalism \cite{NP}. It is an equivalent way of writing Einstein field equations. With it we can treat outgoing gravitational waves around massive objects like black holes, neutron stars or wormholes.

This spin-coefficient version is a variation on Ricci's method where the orthonormal tetrad is replaced by null vectors or two-component spinors. Its relevance to outgoing gravitational radiation is contained in the Weyl complex valued scalar field $\Psi_4$ and has been verified to coincide with the linearized gravitation prediction for empty space. This formalism was used to produce a linear equation for the perturbation of a rotating black hole (Kerr metric), the Teukolsky equation, which in its original form contains the stress-energy tensor of the material content \cite{TE}. However, the treatment of the material content is not self-consistent, that is the equation is written for the empty space background metric (type D -- Kerr or Schwatzschild), not for the metric created by the particular material content. It serves as a perturbation and therefore the Teukolsky equation can be viewed as a generalization to the linearized gravitation for black hole geometries. Its usefulness for the exploration of the stability of empty space solutions is irreplaceable.

Note, in what follows we will remain within the einsteinean coordinate-basis version using the metric tensor components as the basic variables and the Christoffel symbols for the connection. Often times in physics it is the framework or choice of coordinates which yields the simplest and clearest description. We would be striving for such a description in addition to self-consistency, that is a gravitational wave treatment around a curved space-time produced by its respective source.


To go beyond the linear approximation, it suffices to assume that the variation to the metric (any general background one) is a small quantity to which we can ascribe the 1st order in the expansion of the equations around the background metric. The resulting equations of motion are the extension to the linear gravitation in the case of a black hole immediate proximity. With such a theory at hand, one can further probe the properties of gravitational waves and their quanta (gravitons) in cosmological setting.

The paper is organized as follows. Section \ref{sec:GW_eqns} introduces the proper mathematical procedure to arrive at the equations for the gravitational waves beyond the linear approximation. Section \ref{sec:MassiveGraviton} explores the consequences of the elementary excitations to the resulting equations being massive particles. Fortunately, important observational facts such as the rotational profile of galaxies and the Lambda Cold Dark Matter ($\Lambda$CDM) cosmological model are recovered. Dark matter comes about as an effect of the non-vanishing graviton mass. Section \ref{sec:Quantization} solves the equation for the gravitational waves in and around the Schwarzschild black hole. The most important observational and experimental consequence of that solution being the ability of black holes to scatter gravitational waves. In the next Section \ref{sec:mirror} we write the equation in the Newtonian limit only to discover the emergence of an index of refraction $n=1-2\Phi/c^2$ which depends on the local gravitational potential $\Phi$ and the velocity of light in vacuum $c$. Interestingly, the local gravitational potential can be manipulated by creating structures of alternating sparse and dense materials such as aluminum and iridium thus opening the possibility of making a gravitational wave mirror and consequently its application in the field propulsion scheme discussed in Section \ref{sec:propulsion}. The paper ends with a Conclusions section. 

\section{Gravitational wave equation beyond the linear approximation}\label{sec:GW_eqns}

\subsection{Varying the Ricci tensor}

Consider the variation of the Ricci tensor from the Palatini identity \cite{palatini}
\begin{eqnarray}\label{Palatini}
	\delta R_{\sigma \nu} = \nabla_{\alpha} \delta \Gamma^{\alpha}_{\; \nu \sigma} -
	\nabla_{\nu} \delta \Gamma^{\alpha}_{\; \alpha \sigma} .   
\end{eqnarray}    
Here $\nabla_{\rho}$ is the covariant derivative and $\delta \Gamma$ is the variation of the Christoffel symbol. Notably, Palatini-like identity holds for the Lie derivative $\mathcal{L}_{\xi} R_{\sigma \nu}$ of the Ricci tensor as well thus confirming the deep geometrical meaning of the identity itself. 

The Palatini identity can be derived by (a) varying the Riemann curvature tensor given by the Levi-Civita connection and (b) making use of the fact that the variation $\delta \Gamma^{\alpha}_{\; \nu \sigma}$ of the Christoffel symbol is a tensor unlike the symbol itself. Upon taking its covariant derivative to substitute for the partial derivatives in step (a) all $\Gamma \delta \Gamma$ terms cancel out and one arrives at 
\begin{eqnarray}\label{delta_Riemann}
	\delta R^{\alpha}_{\; \sigma \mu \nu} = \nabla_{\mu} \delta \Gamma^{\alpha}_{\; \nu \sigma} -
	\nabla_{\nu} \delta \Gamma^{\alpha}_{\; \mu \sigma}.    
\end{eqnarray} 
Contracting $\alpha$ and $\mu$ in (\ref{delta_Riemann}) produces (\ref{Palatini}): $\delta R_{\sigma \nu}=\delta R^{\alpha}_{\; \sigma \alpha \nu}$.

Next, we focus on substituting the variation of the Christoffel symbol
\begin{eqnarray}\label{delta_Christoffel}
	\delta \Gamma^{\beta}_{\; \nu \sigma} = \frac12 g^{\beta \alpha} \left(
	\nabla_{\nu} \delta g_{\alpha \sigma} + \nabla_{\sigma} \delta g_{\alpha \nu} - \nabla_{\alpha} \delta g_{\nu \sigma}     
	\right)	    
\end{eqnarray} 
into the Palatini identity to obtain in a straightforward calculation (alternatively going through Riemann normal coordinates) an expression for the variation of the Ricci tensor
\begin{eqnarray}\label{delta_Ricci}
	\delta R_{\sigma \nu} = - \frac12 \left(    
	\nabla_{\sigma} \nabla_{\nu} g^{\alpha \beta} \delta g_{\alpha \beta} + \nabla^{\gamma} \nabla_{\gamma} \delta g_{\sigma \nu}
	\right).   
\end{eqnarray}    
Contracting this result with the metric tensor yields an even simpler form 
\begin{eqnarray}\label{delta_Ricci_contracted}
	g^{\sigma \nu}\delta R_{\sigma \nu} = - \Box \;  g^{\alpha \beta} \delta g_{\alpha \beta}  
\end{eqnarray}    
by virtue of $\nabla_{\gamma} g_{\sigma \nu}=0.$ Here $\Box = \nabla^{\gamma} \nabla_{\gamma}$ is the d'Alambert wave operator given in terms of the metric and its determinant $g$ by
\begin{eqnarray}\label{dAlambert}
	\Box \; \psi =  \frac{1}{\sqrt{-g}} \frac{\partial}{\partial x_\alpha} \left( 
	\sqrt{-g} g^{\alpha \beta} \frac{\partial \psi}{\partial x_\beta} 
	\right).
\end{eqnarray}

\subsection{Varying the material content tensor}

The stress-energy tensor is defined as the functional derivative of the matter action $S_m$ with respect to the metric
$
T_{\mu\nu} = -\frac{2}{\sqrt{-g}} \frac{\delta S_{m}}{\delta g^{\mu\nu}}.
$
However, we need the {\it variation} of the material content tensor with respect to the metric, i.e. $\delta M_{\sigma\nu} $, to construct the new equation describing the metric perturbations:
\begin{eqnarray}\label{MC_variation}
	\delta M_{\sigma \nu} &=& \left(\Lambda - \frac{1}{2}\kappa T  \right)\delta g_{\sigma \nu} + \kappa \left( \delta T_{\sigma \nu} - \frac12 g_{\sigma \nu} \delta T\right).   
\end{eqnarray}  
Here we have assumed that the cosmological term ${\delta \Lambda}/{\delta g^{\mu\nu}}=0$ does not depend on the metric. 

The question ultimately is about how $T_{\mu\nu}$ and its trace $T$ change when the metric is perturbed. To compute this, we need to consider the explicit dependence of  $T_{\mu\nu}$ on the metric, which depends on the specific form of the matter content (e.g., scalar field, electromagnetic field, perfect fluid).

Naturally, we want to start with the simplest example, that is a perfect fluid
\begin{eqnarray}\label{dust}
	T_{\sigma \nu} = \left(\rho + \frac{p}{c^2} \right) u_{\sigma} u_{\nu} + g_{\sigma \nu} p,   
\end{eqnarray}    
where $\rho$ is the mass-energy density, $u_{\sigma}$ is the four-velocity of the particles with the property $u^{\alpha} u_{\alpha}=-c^2$ and $p$ is the hydrostatic pressure exerted by the gas of particles. The trace yields $T=3p-\rho c^2.$

Under a small variation $g_{\mu\nu} \to g_{\mu\nu} + \delta g_{\mu\nu} $, the variation of $T_{\mu\nu}$ is:
\begin{eqnarray}\label{delta_T mu nu}
\delta	T_{\sigma \nu} = \delta \left(\rho + \frac{p}{c^2} \right) u_{\sigma} u_{\nu} + \left(\rho + \frac{p}{c^2} \right) \delta (u_{\sigma} u_{\nu}) + g_{\sigma \nu} \delta p  + p \delta g_{\sigma \nu}  
\end{eqnarray} 
which is not simple enough and we will further simplify by assuming dust as a specific type of perfect fluid used to model matter that has negligible pressure $p=0$ compared to its energy density. This is a common approximation for non-relativistic matter, such as a collection of collisionless particles (e.g., galaxies in cosmology or cold dark matter). In this case (\ref{delta_T mu nu}) becomes
\begin{eqnarray}\label{delta_T dust}
	\delta	T_{\sigma \nu} =  \rho u_{\sigma} u^{\mu} \delta g_{\mu \nu} + \rho  u_{\nu} u^{\mu } \delta g_{\mu \sigma} 
\end{eqnarray}
and
\begin{eqnarray}\label{contracted_delta_T dust}
	g^{\sigma \nu} \delta	T_{\sigma \nu} =   2 \rho  u^{\sigma} u^{\mu }  \delta g_{\mu \sigma}=0.  
\end{eqnarray}
Here we can drop the term $2 u^{\nu} u^{\mu} \delta g_{\mu \nu}$ due to it being a difference of the same constant: $u^{\nu} u^{\mu} g'_{\mu \nu} - u^{\nu} u^{\mu} g_{\mu \nu}=-c^2+c^2=0$.

The variation of the trace is the other object of value, that is
\begin{eqnarray}\label{delta_trace}
	\delta	T = \delta \left( 3p - \rho c^2  \right) = 0,  
\end{eqnarray}
since the equation of state $p(\rho)$ for a perfect fluid does not depend on the metric in the usual formulation of relativity. It is a fundamental property of the fluid, independent of the space-time geometry.

The ultra-relativistic gas case is not much different. Substantially more interesting is the electromagnetic field case. For it
\begin{eqnarray}\label{T_EMfield}
	T_{\sigma \nu}^{EM} = \frac{1}{\mu_0} \left( F_{\sigma \alpha} F_{\nu}^{\; \alpha} - \frac14 g_{\sigma \nu} F_{\alpha \beta} F^{\alpha \beta}\right)    
\end{eqnarray}
and its variation
\begin{eqnarray}\label{T_EMfield_variation}
	\delta T_{\sigma \nu}^{EM} = \frac{1}{\mu_0} \left(- F^{\; \alpha}_{\sigma} F_{\nu}^{\;\beta} \delta g_{\alpha \beta} - \frac14  F_{\alpha \beta} F^{\alpha \beta} \delta g_{\sigma \nu} + \frac12 g_{\sigma \nu} F_{\alpha \beta} F^{\alpha \beta} \delta g^{\alpha \beta}   \right).    
\end{eqnarray}
Contracting this quantity with the metric tensor yields
\begin{eqnarray}\label{contracted_T_EMfield_variation}
	g^{\sigma \nu}\delta T_{\sigma \nu}^{EM} = \frac{1}{\mu_0}  \left(- F^{\nu \alpha} F_{\nu}^{\;\beta} \delta g_{\alpha \beta} - \frac14  F_{\alpha \beta} F^{\alpha \beta} g^{\sigma \nu}\delta g_{\sigma \nu} + 2 F_{\alpha \beta} F^{\alpha \beta} \delta g^{\alpha \beta}   \right).    
\end{eqnarray}
Note that the variation of the vanishing trace $T^{EM}=0$ is also vanishing $\delta T^{EM}=0$.

\subsection{ Notion of gauge invariance}

When contracting the variation of the  material content tensor (\ref{MC_tensor}) or the variation of the Ricci tensor (\ref{delta_Ricci_contracted}) with the metric tensor $g^{\sigma \nu} \delta M_{\sigma \nu}$ and $g^{\sigma \nu}\delta R_{\sigma \nu}$, we obtain a scalar field 
\begin{eqnarray}\label{GW_amplitude}
	\psi= g^{\sigma \nu}\delta g_{\sigma \nu} 	\,,
\end{eqnarray}
which acts like a generalized amplitude of the gravitational wave and is central to our discussion. It is closely related to the trace of the metric perturbation $g^{\sigma \nu}\delta g_{\sigma \nu} \to \eta^{\sigma \nu}h_{\sigma \nu}=h_{\sigma}^{\sigma}$ in the linearized version of gravitation $g_{\sigma \nu}=\eta_{\sigma \nu}+h_{\sigma \nu}$ ($| h_{\sigma \nu} | \ll 1$ and $\eta_{\sigma \nu}={\rm diag} (-1,1,1,1)$) and therefore subject to controversy stemming from the famous transverse-traceless (TT) gauge in the description of linearized waves.

The vanishing of the trace is achieved by exploiting the gauge invariance of linearized gravity. Under an infinitesimal coordinate transformation \( x^\mu \to x^\mu + \xi^\mu \), the perturbation transforms as:
\[ h_{\mu\nu} \to h_{\mu\nu} - \partial_\mu \xi_\nu - \partial_\nu \xi_\mu. \]
The TT gauge is a specific choice of \( \xi^\mu \) that enforces the transverse and traceless conditions. The traceless conditions demand $\partial_\nu \xi^\nu=0$ a vanishing divergence.

How does the quantity $\psi=g^{\sigma \nu}\delta g_{\sigma \nu}$ transform under the same coordinate transformations $x'^{\mu}=x^{\mu}+\xi^\mu$? We argue it transforms like a scalar field. The change in the metric due to the coordinate shift is given by the Lie derivative along $\xi^{\mu}$. Here $\nabla$ is the covariant derivative: $\mathcal{L}_{\xi} g_{\sigma \nu}=-\nabla_{\sigma}\xi_{\nu}-\nabla_{\nu}\xi_{\sigma}$. The inverse metric transforms inversely: $\mathcal{L}_{\xi} g^{\sigma \nu}=\nabla^{\sigma}\xi^{\nu}+\nabla^{\nu}\xi^{\sigma}$. The variation of the metric $\delta g_{\sigma \nu}$ is distinct from the Lie derivative $\mathcal{L}_{\xi} g_{\sigma \nu}$. Therefore $\delta g_{\sigma \nu}$ transforms like a tensor: $ \mathcal{L}_{\xi} (\delta g_{\sigma \nu})= -(\nabla_{\sigma} \xi^{\alpha}) \delta g_{\alpha \nu} - (\nabla_{\nu} \xi^{\beta}) \delta g_{\beta \sigma} $. Finally, $\psi$ transforms as $\psi'=g'^{\sigma \nu}\delta g'_{\sigma \nu}=[g^{\sigma \nu} + \mathcal{L}_{\xi} g^{\sigma \nu} ] [\delta g_{\sigma \nu}+ \mathcal{L}_{\xi} (\delta g_{\sigma \nu})]$. Alternatively, $\delta \psi = \psi'- \psi = (\mathcal{L}_{\xi} g^{\sigma \nu} )\delta g_{\sigma \nu} + g^{\sigma \nu} \mathcal{L}_{\xi} (\delta g_{\sigma \nu}) + o(\xi^2) = (\nabla^{\sigma} \xi^{\nu} + \nabla^{\nu} \xi^{\sigma})\delta g_{\sigma \nu} - [\nabla^{\nu} \xi^{\alpha} \delta g_{\alpha \nu} + \nabla^{\sigma} \xi^{\beta}\delta g_{\sigma \beta}] =0$. (End of proof.) 

Indeed, the scalar field $\psi=g^{\sigma \nu}\delta g_{\sigma \nu}$ is invariant under coordinate transformations. It cannot be made to vanish by the choice of the gauge condition. Therefore, the ``traceless gauge'' from the linearized theory is not applicable in the general case discussed here. It is often claimed that the traceless gauge removes scalar perturbations (e.g., breathing modes), leaving only the tensor modes, which are considered  the ``true'' physical degrees of freedom of gravitational waves (2 polarizations in 4D spacetime).  This can work in the linear regime where there is no distinction between covariant and partial derivatives. Its equivalent in the general case discussed here is a vanishing divergence for the coordinate shift: $\partial_\nu \xi^\nu \to \nabla_\nu \xi^{\nu}=0$.

\subsection{The tensor equation}

Let us equate (\ref{delta_Ricci}) with (\ref{MC_variation}) to obtain the tensor equation $\delta R_{\sigma \nu} = \delta M_{\sigma \nu}$, 
\begin{eqnarray}\label{Tensor equation}
	- \frac12 \left(    
	\nabla_{\sigma} \nabla_{\nu} g^{\alpha \beta} \delta g_{\alpha \beta} + \nabla^{\gamma} \nabla_{\gamma} \delta g_{\sigma \nu}
	\right) &=& \mathcal{M} \delta g_{\sigma \nu} + \mathcal{J}_{\sigma \nu} \,,
\end{eqnarray} 
where the field variable is the amplitude of the perturbation $\delta g_{\sigma \nu}$, which is exactly what we call gravitational waves and is essentially the difference of two almost equal metric tensors, therefore a tensor itself. Here 
\begin{eqnarray}\label{M}
	\mathcal{M} &=&  \Lambda - \frac{1}{2}\kappa T  , \\
	\mathcal{J}_{\sigma \nu} &=& \kappa \left( \delta T_{\sigma \nu}  - \frac12 g_{\sigma \nu} \delta T\right). 
\end{eqnarray} 
The Lorentz covariance of (\ref*{Tensor equation}) is obvious and these are the equations of motion for the gravitational waves beyond the linear approximation. They describe the real physical dynamics by virtue of them being a ``difference" in the Ricci tensor (equivalent form of Einstein field equation) for two almost equal space-times.  However, they are not straightforward to solve or interpret. We are going to restrict the discussion in this paper to the contracted quantity, that is the scalar field $\psi$ from (\ref{GW_amplitude}). The eliminated information allows us to present a noteworthy equation.

A curious feature of this equation is the emergence of a term which is reminiscent of the mass term in a relativistic field theory. It is the $\mathcal{M}$ term that has the structure {\it mass$^2$ $\times$  field}. Typically in relativistic theories the coefficient multiplying the field variable is considered the squared mass of the field's particle excitations, which in this case are the gravitons. Notably, for certain types of material content, like the electromagnetic field, $T=0$ and $m^2=\mathcal{M}=\Lambda\approx 1.1 \times 10^{-52} {\rm m}^{-2} > 0$ (from Hubble tension measurements). Here we can identify the purpose of the cosmological term -- it fixes the mass of the gravitational wave excitations and keeps it real-valued, thus allowing for stable, oscillatory and propagating waves. 

However, for matter content (perfect fluid $T = 3p - \rho c^2 $) of sufficient pressure $p> \rho c^2 + 2 \Lambda /\kappa$, the effective mass can become imaginary:  $m^2=\mathcal{M}< 0$ by virtue of the trace of the stress-energy tensor overcoming the cosmological term. The consequences for these regimes in the equation remain a work in progress but the tachyon-like, in the context of quantum field theory, interpretation is very plausible (and associated instabilities like collapse or exponential growth of the field variable, that is the gravitational field).

Exotic forms of matter such that in their equation of state $p(\rho)$ the pressure is bigger than the energy density might be found beyond the Zel’dovich fluid, that is the causal limit. In certain quantum field theories or condensed matter systems (e.g., strongly interacting fermions), effective equations of state might yield $p> \rho c^2$ locally due to interactions or negative energy contributions. We have identified possible candidates in a previous publication \cite{VS}. In such a context, the gravitational wave generated by, say an oscillatory quadrupole configuration, may become unstable and thus substantially modify its space-time embedding. An experimental verification of such a prospect would be a worthy effort.

Naturally, the last term on the right hand side, that is $\mathcal{J}_{\sigma \nu}$, does not explicitly contain the field variable and is in essence the variation of the stress-energy tensor and its trace. We recognize its presence as the source of the gravitational waves described by the relativistic part on the left hand side. Coincidently, the gravitational wave encoded in the variation of the metric is sourced from the variation of the stress-energy tensor. The stress-energy tensor produces the background metric on top of it live the small perturbations we call gravitational waves. Their source must therefore be the perturbation to the source of the background metric. Note, this feature is absent in the linearized theory of gravitational waves. There one finds that the perturbations to the metric are proportional to the stress-energy tensor itself. Such a result can only lead to false predictions which are circumvented in literature by  virtue of considering very weak sources or abandoning the material content altogether. Most of the results in the gravitational wave physics within the linearized theory are produced in the TT gauge, which is admissible only for pure waves and not for more general solutions with a source (see p. 947 and p. 949 of \cite{MTW}).

\subsection{Going beyond the linear approximation}

Contracting (\ref*{Tensor equation}) with the inverse metric tensor $g^{\sigma \nu}$ produces
\begin{eqnarray}\label{GW_eqn}
	 \Box \;  \psi  + m^{2} \psi &=& \mathcal{J}, 
\end{eqnarray}    
where 
\begin{eqnarray}\label{source}
	\mathcal{J} &=& - g^{\sigma \nu} \mathcal{J}_{\sigma \nu}  = \kappa \left( 2 \delta T - g^{\sigma \nu} \delta T_{\sigma \nu}  \right), 
\end{eqnarray} 
serves as the source of the gravitational radiation,
where the ``effective mass'' is given by
\begin{eqnarray}\label{GW_mass}
	m^{2} &=& \mathcal{M} = \Lambda - \frac{1}{2}\kappa T   
\end{eqnarray}
and defined as a square root of (\ref{M}).  The generalized amplitude of the gravitational wave is defined in (\ref{GW_amplitude}) as $ \psi =  g^{\sigma \nu} \delta g_{\sigma \nu}$. The connection with the mass of a particle $m_g$ described by such an equation goes through a relation of the sort $m_g = {m\hbar}/{c}$. 

Note, the source term is proportional to the variation of the stress-energy tensor and its trace multiplied with the Einstein constant thus rendering it exceedingly small. In effect, the source term can be set to zero to obtain 
\begin{eqnarray}\label{GW_eqn_j0}
	\Box \;  \psi  + m^{2} \psi & = & 0 \,, 
\end{eqnarray} 
which is a free massive Klein-Gordon equation for the generalized amplitude $\psi$ of the gravitational wave. Note the mass term cannot be set to zero, due to the cosmological term fixing its value in empty space. We are tempted to speculate, that this might be the reason for the existence of dark energy (cosmological term). Something in the Universe must maintain gravitational wave stability (real-valued mass) even in the absence of matter.

Note, the d'Alambert operator (\ref{dAlambert}) is defined on curved space-time and the resulting Klein-Gordon equation is a relativistic wave equation \cite{KG1, KG2, KG3, KG4, KG5, KG6, KG7}. In the case of gravitational waves it has an associated effective mass and is manifestly Lorentz covariant. There is no interaction term present for the particulate matter. This might not be the case for electromagnetic fields and remains a research topic in progress. 

The solution for the scalar field $\psi$ typically describes spin-less particles. The Higgs boson \cite{higgs1, higgs2, higgs3} (a spin-zero particle) is the first observed relevant elementary particle to be described by the Klein–Gordon equation \cite{cern1, cern2}.

In effect, any component of any solution to the free Dirac equation is also a solution to the free Klein–Gordon equation. Due to the Bargmann–Wigner equations, the result generalizes to particles of arbitrary spin \cite{BW}. Therefore, in quantum field theory any component of any quantum field should satisfy the free Klein–Gordon equation. There cannot be a more generic expression of quantum fields' dynamics.

Since this gravitational wave theory is massive in essence, an important note regarding the Weinberg-Witten theorem \cite{WW1} is to be made. The theorem states that a four dimensional Lorentz-invariant quantum field theory with a conserved, gauge-invariant stress tensor cannot
have massless particles with spin $>1$. Indeed, our graviton, that is the particle mode of (\ref{GW_eqn}) comes out massive and the equation has all the covariance one needs, which is a confirmation that the description is in line with the Weinberg-Witten theorem \cite{WW1}.

\section{Massive Graviton}\label{sec:MassiveGraviton}

The mechanisms through which the graviton, that is the particle assumed to mediate the gravitational interaction, can acquire mass are different and include processes such as  (i) Higgs-like spontaneous symmetry breaking induced by condensation of scalar fields \cite{CMuk} or (ii) modified or extended theories of gravitation exhibiting Yukawa gravitational potential as a correction to the Newton's potential. The Yukawa potential appears in $f(R)$ theories of gravitation \cite{K2, C1, C2, C3, C4} and leads to the emergence of an effective $\Lambda$CDM model.

These ideas deviate from the usual weak-field limit interpretation of the Einstein's theory, where the graviton is massless and therefore propagates with the velocity of light in vacuum. This has led to insufficient exploration of massive theories such as Refs. \cite{MT1, MT2, MT3, MT4, MT5, MT6, MT7}.

As we have demonstrated in the previous sections, going beyond the linear approximation in the Einstein theory produces equations of motion clearly exhibiting massive gravitons. The derived equation (\ref{GW_eqn}) is also strikingly similar to the equation of massive gravitation proposed by Fierz and Pauli \cite{MT1}.

Graviton's mass should be a function of the surrounding matter and energy (including dark) content, which is consistent with the idea that graviton's mass can fluctuate on cosmological scales \cite{K2, K1} and is the case in theories of massive gravity \cite{M1, M2, M3}. Note, the results of NANOGrav collaboration show potential contribution of massive gravitons to the gravitational wave spectrum in the nHz range \cite{NG, Wu}. Moreover, NANOGrav 15-year data release contains the first evidence for a stochastic gravitational wave background which may be considered as a proof of the existence of gravitational waves beyond the results of LIGO. In particular, it included the first measurement of the Hellings-Downs curve \cite{HD}, the tell-tale sign of the gravitational wave origin of the observed background.

The expected graviton mass $m_g$ is exceedingly small (\ref{GW_mass}) and in the present work it is given by
\begin{eqnarray}\label{GW_mass_II}
	m_g = \frac{\hbar}{c} \sqrt{ \Lambda - \frac{1}{2}\kappa T } .  
\end{eqnarray}   
As a result, it is necessary to have a quantum mechanical description (quantum gravity) which is presently non-existent. However, the Heisenberg's uncertainty principle $\Delta p \Delta x \sim \hbar$ must hold with the effect of having different uncertainties in momentum, ergo mass, on stellar, galactic and cosmic scales (due to different uncertainties in position at different scales). An experimental consequence is the existence of minimal frequency in the gravitational wave spectrum (akin to \cite{Wu})
\begin{eqnarray}\label{GW_freq_min}
	\nu_{min}=\frac{m_g c^2}{2 \pi \hbar} = \frac{c}{2 \pi} \sqrt{\Lambda - \frac{1}{2}\kappa T } .  
\end{eqnarray}

The cosmological consequences of massive gravitons were reported in Refs. \cite{CC1, CC2, CC3} and are compelling since they recover the Modified Newtonian Dynamics (MOND) necessary to explain the observed rotation curves of galaxies without the introduction of the notion of dark matter. Dark matter manifests itself as an apparent effect. Furthermore, $\Lambda$CDM cosmological model is also recovered. Such a resolution of the nature of dark matter is much needed, since as it stands now dark matter is a ``mysterious'' form of matter that interacts only gravitationally \cite{DM1, DM2, DM3, DM4, DM5}. As a result, there has been no direct detection of its quantum excitations, that is, dark matter particles. The recovery of the $\Lambda$CDM cosmological model is promising since this model accounts for a wide range of cosmological observations at the expense of a minimal set of parameters \cite{CDM}.

Massive gravitons in modified theories of gravity are associated with complications \cite{DB} which ultimately lead to instabilities. Later versions of such theories worked out a compromise \cite{IN1, IN2, IN3}. In the present framework, no modifications to the Einstein field equations have been made. The emergence of a massive graviton takes place only in the presence of matter or dark energy via the cosmological term $\Lambda$. We, therefore, do not expect the emergence of any instability, where the resolution of this issue is beyond the scope of the present paper.

\section{Quantization}\label{sec:Quantization}

A reasonable first step would be to consider (\ref{GW_eqn}) on the Schwarzschild metric, the quintessential black hole line element that has the form
\begin{eqnarray}\label{Schwarzschild metric}
	ds^2=-F(r)c^2 dt^2 + F^{-1} dr^2 + r^2 d\Omega^2,	   
\end{eqnarray}    
where $c$ is the speed of light in vacuum, $\Omega$ is a point on the two-sphere $\mathbb{S}^2$ and $F(r)=1-r_s/r$. Here the Schwarzschild radius is given by $r_s=2GM/c^2$, where $G$ is the Newton's gravitational constant and $M$ is the mass of the black hole or the body creating the curvature in the space-time canvas. Note for $r>r_s$ the time coordinate is given by $t$ and signifies the measurement by a clock located at infinity with respect to the massive body. Usually, $d\Omega^2=d \theta^2 + \sin^2 \theta d\phi^2$, where $\theta$ is the latitude of $\Omega$ after arbitrarily choosing a $z$-axis and $\phi$ is the longitude of $\Omega$ around the chosen $z$-axis.

After a lengthy but straightforward calculation involving the metric tensor $g_{\mu \nu}={\rm diag}\left( -F, F^{-1}, r^2, r^2\sin^2\theta \right)$ and its inverse, for (\ref{GW_eqn}) we obtain
\begin{eqnarray}\label{eq:GW_Schwarzschild}
	\left[-F^{-1} \partial^2_{ct}	+ \frac{1}{r^2} \partial_r \left( r^2 F \partial_r \right) + \frac{1}{r^2} \Delta_{\mathbb{S}^2} +  \mu^{2}
	\right]\psi=0.
\end{eqnarray} 
We multiply the equation with $r_s^2$ to introduce the dimensionless variable $x=r/r_s$ which measures radial distance in terms of units of Schwarzschild radius
\begin{eqnarray}\label{eq:x_GW_Schwarzschild}
	\left[-\frac{r_s^2}{F(x)} \; \partial^2_{ct}	+ \frac{1}{x^2} \partial_x \left( x^2 F \partial_x \right) + \frac{1}{x^2} \Delta_{\mathbb{S}^2} +  r_s^2\mu^{2}
	\right]\psi=0.
\end{eqnarray} 
In terms of the new dimensionless variables, the singularity remains at $x=0$ and the event horizon is at $x=1$. Here
\begin{eqnarray}\label{F(x)}
	F(x)=1-\frac{1}{x}.
\end{eqnarray}

Subsequently, we split the dependence on the variables in the solution
\begin{eqnarray}\label{}
	\psi(t,r,\theta, \phi)= e^{i \omega t}\,  \Psi(x)\, Y^{m}_{l}(\theta, \phi) 
\end{eqnarray} 
and divide by $\psi$ to obtain 
	\begin{eqnarray}\label{eqns:split}
		-\frac{r_s^2}{F(x)} \; \frac{1}{e^{i \omega t}}\partial^2_{ct} e^{i \omega t} = \frac{r_s^2 }{ F(x)} \frac{\omega^2}{c^2} = A, \\
		\frac{1}{x^2} \frac{1}{Y^{m}_{l}}\Delta_{\mathbb{S}^2} Y^{m}_{l} = -\frac{l(l+1)}{x^2}=B, \\
		\left[\frac{1}{x^2} \frac{d}{dx} \left( x^2 F \frac{d}{dx} \right) + A + B +  r_s^2\mu^{2}
		\right]\Psi(x)=0.
	\end{eqnarray} 

Here $Y^{m}_{l}(\theta, \phi)$ is a spherical harmonic function of degree $l$ and order $m$ ($l\in \mathbb{N}$ and $m \in \mathbb{Z}$), which are solutions to \cite{Math1, Math2}
\begin{eqnarray}\label{}
	\Delta_{\mathbb{S}^2} Y^{m}_{l}(\theta, \phi) = - l(l+1) \; Y^{m}_{l}(\theta, \phi)
\end{eqnarray} 
and are typically given in terms of the associated Legendre polynomial $P^{m}_{l}(\cos\theta)$ and a normalization constant $C$ as $
Y^{m}_{l}(\theta, \phi)=C e^{i m \phi} P^{m}_{l}(\cos\theta).$
There are $2l+1$ such polynomials and $m$ runs from $-l$ to $l$ in integer steps. Since the solution to the equation signifies the perturbation to the metric in terms of gravitational waves we need a real valued solution, therefore $e^{i m \phi}$ is to be replaced with  $\sin(m \phi)$ or $\cos(m \phi)$.

Now let us focus on the radial part
\begin{eqnarray}\label{eqn: Psi}
	\left[\frac{1}{x^2} \partial_x \left( x^2 F \partial_x \right) + \frac{r_s^2 }{ F(x)} \frac{\omega^2}{c^2}  -\frac{l(l+1)}{x^2} +  r_s^2\mu^{2}
	\right]\Psi(x)=0
\end{eqnarray} 
and more specifically on the differential operator part
\begin{eqnarray}\label{eqn:Psi_diff}
	\left[\frac{1}{x^2} \frac{d}{dx} \left( x^2 F \frac{d}{dx} \right)\right]\Psi(x)= F(x) \frac{d^2 \Psi}{dx^2} + \left( \frac{dF}{dx} + \frac{2F}{x}  \right) \frac{d\Psi}{dx}.
\end{eqnarray} 
The presence of the first derivative of $\Psi$ is highly undesirable and we can remove it by the ansatz 
\begin{eqnarray}\label{eqn:ansatz}
	\Psi(x)= N e^{-\ln(x\sqrt{F}\,)}\chi(x),
\end{eqnarray} 
where $N$ is a normalization constant. As a result (\ref{eqn:Psi_diff}) becomes 
\begin{eqnarray}\label{}
	\nonumber	\left[\frac{1}{x^2} \frac{d}{dx} \left( x^2 F \frac{d}{dx} \right)\right]\Psi(x)= N e^{-\ln(x\sqrt{F}\,)} F(x) \left( \frac{d^2 \chi}{dx^2}  + \frac{1}{4}\frac{1}{x^2(x-1)^2} \, \chi  \right).
\end{eqnarray} 
Finally, for the radial part (\ref{eqn: Psi}) we obtain {\it an effective non-relativistic Schr\"odinger equation}
\begin{eqnarray}\label{eqn: chi}
	\left[-\frac{d^2 }{dx^2}  + V_{eff}(l, x)
	\right]\chi(x)=0,
\end{eqnarray} 
where the effective potential is given by
\begin{eqnarray}\label{V_eff}
		V_{eff} (l, x)&=& - \frac{1}{4}\frac{1}{x^2(x-1)^2} + \frac{l(l+1)}{x^2 F(x)} - \frac{r_s^2 }{ F^2(x)} \frac{\omega^2}{c^2} -  \frac{r_s^2\mu^{2}}{F(x)} \\
\nonumber	&=& - \frac{1}{4}\frac{1}{x^2(x-1)^2} + \frac{l(l+1)}{x (x-1)} - \frac{ x^2}{ (x-1)^2} \frac{r_s^2 \omega^2}{c^2} -  r_s^2\mu^{2} \frac{ x}{(x-1)} ,
\end{eqnarray}
with the following behavior at the singular points
	\begin{eqnarray}\label{}
\nonumber		V_{eff} |_{x \to 0} & \approx& - \frac{1}{4 x^2} ,  \\
\nonumber 		V_{eff} |_{x \to 1} & \approx&  - \frac{ 1}{ (x-1)^2}\; \left[\frac{r_s^2 \omega^2 }{c^2} + \frac14 \right], \\
\nonumber		V_{eff} |_{x \to \infty} & \approx&  -  \left[\frac{r_s^2 \omega^2}{c^2} +  r_s^2\mu^{2}\right].
	\end{eqnarray} 

Let us briefly discuss the properties of the effective potential. For
$l=0$ orbital momentum state, the effective potential is always negative $V_{eff}(l=0,x)<0$ for all $x \geq 0$. Therefore, the solutions are bound states, i.e. standing gravitational waves trapped in the interior of the black hole.  For non-vanishing orbital momentum states $l\geq 1$ the effective potential preserves its shape in the interior of the black hole. However, in the exterior part $x \geq 1$, it is no longer negative but becomes repulsive (positive). There is a marked maximum right after the Schwarzschild radius ($dV_{eff}/dx=0$ has a solution, $x_l>1$). The height of the maximum rises rapidly with the increase of the orbital momentum of the state. As a result, there are strong conditions for gravitational wave reflection before the Schwarzschild radius is reached coming in from infinity. This provides an experimental condition for the verification of the existence of black holes as described by the Einstein field equations.

\begin{figure}[t]
	\begin{center}
		\includegraphics[scale=0.4]{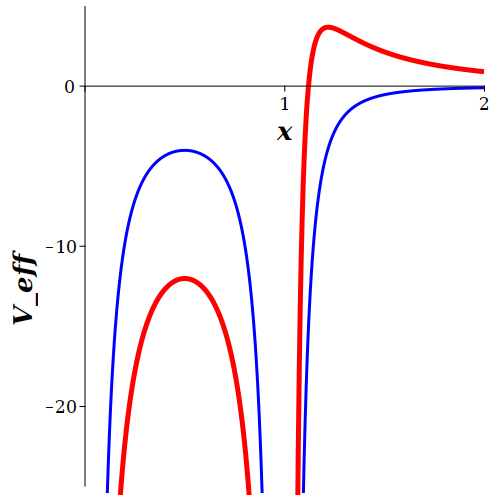}
		\caption{\label{fig1} The effective potential for the radial part of the equation for gravitational waves on Schwarzschild metric. The blue line corresponds to $l=0$ orbital momentum and is strictly negative (bound states). The red line corresponds to $l=1$ orbital momentum. The positive energy peak for $l \geq 1$ states right after the event horizon at $x=1$ points to gravitational waves coming in towards the singularity at $x=0$ from infinity being scattered off the event horizon. Here ${r_s^2 \omega^2}/{c^2}=0.01$ and $r_s^2\mu^{2}=0.0001$.
		}
	\end{center}
\end{figure}

Presently, the existence of black holes is only inferred by the orbital motion of a companion star around an invisible very massive center \cite{EH1}. In this paper we provide a very concrete theoretical prediction towards an unambiguous proof of the existence of black holes. Suppose a gravitational wave traveling towards the black hole has energy less than the height of the potential barrier depicted in Figure \ref{fig1}. Since the barrier is higher, there would be a reflected wave besides the transmitted one. Therefore, an observation of gravitational wave reflection from a potential black hole candidate could serve as an unambiguous indication of the existence of black holes.

\begin{figure}[t]
	\begin{center}
		\includegraphics[scale=0.4]{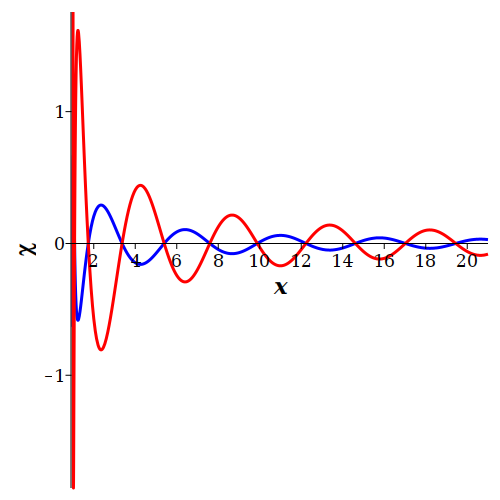}
		\caption{\label{fig2} The solution to the radial part (\ref{eqn: Psi}) as dictated by the ansatz (\ref{eqn:ansatz}) in the exterior region $x \geq 1$. The blue line is the real part and the red line is the imaginary part. Note the typical oscillatory behavior for large $x$. Such a behavior is expected due to $V_{eff}$ (\ref{V_eff}). Here $r_s^2\mu^{2}=0.0001$, ${r_s^2 \omega^2}/{c^2}=0.5 \sqrt{-2r_s^2\mu^{2} + 2l^2 + 2l + 2}$ and $l=1$.
		}
	\end{center}
\end{figure}

\begin{figure}[t]
	\begin{center}
		\includegraphics[scale=0.4]{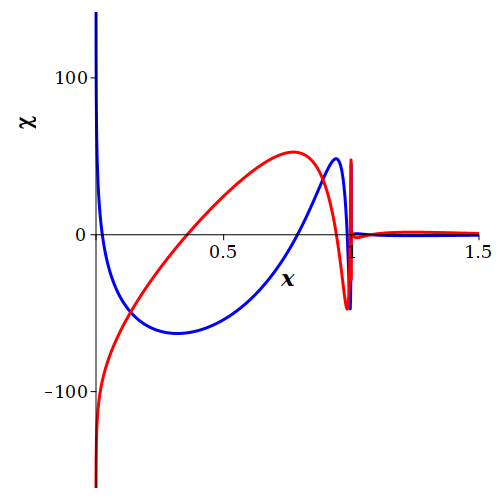}
		\caption{\label{fig3} The solution to the radial part (\ref{eqn: Psi}) as dictated by the ansatz (\ref{eqn:ansatz}) in the interior of the black hole. The blue line is the real part and the red line is the imaginary part. Since inside the black hole the signature of the metric changes, we are no longer certain of the interpretation of (\ref{eqn: Psi}) as the radial part of the equation of motion. Here the parameters are the same as those in Figure \ref{fig2}.
		}
	\end{center}
\end{figure}

There is an analytical solution to the quantum mechanical problem posed in equation (\ref{eqn: chi}) with the potential (\ref{V_eff}). It is by no means trivial and is given in terms of the confluent Heun function. Both the real and imaginary parts of the solution are admissible since (\ref{eqn: chi}) is linear and one can form an appropriate linear combination to reduce the solution to either the real or imaginary part. For the depiction of the solutions, see Figure \ref{fig2} and Figure \ref{fig3}. Most notable is the oscillatory behavior in the exterior to the black hole region. There is no blow up in the solution at the event horizon.

At present we are unable to introduce proper initial or boundary conditions either on the singularity or the event horizon, conditions which would lead to the quantization of the energy spectrum. This remains a research topic in progress.

\section{Gravitational Wave reflection}\label{sec:mirror}

While gravitational waves propagate through the universe, the issue of their reflection is a topic of an ongoing theoretical debate \cite{feynman, weinberg}.  According to the weak field limit in General Relativity, gravitational waves travel along null geodesics, similar to how light rays behave. This means that while gravitational waves can be deflected by massive objects (analogous to gravitational lensing), they do not reflect in the traditional sense as a result of meeting an interface between media where their velocity of propagation is different. The idea of a ``gravitational mirror'' is non-existent in current understanding, as reflection might require a mechanism analogous to the interaction of light (electric field) with matter (electrons), which does not exist for gravitational waves or we have not found a physical set up where such a process may take place. 

Alternatively, matter opposes (due to inertia) the compression and expansion of space-time going on around it as the gravitational wave passes and, as Feynman pointed out, gravitational waves carry energy \cite{feynman}. Matter's kinetic energy increases due to acceleration (absorbs energy from the wave). Dissipation inside the material causes
the energy extracted from the wave to be converted either into heat or re-radiated as a scattered gravitational wave \cite{weinberg}.

A proposal that thin superconducting films might act as mirrors for gravitational waves at microwave frequencies exists \cite{RC1, RC2}. The proposal is based on a decomposition of linearized Einstein field equations, where the transverse-traceless part of the metric perturbation describes gravitational waves in matter. Gravitational waves incident on a superconductor can be described by a linear London-like constituent equation characterized by a “gravitational shear modulus”, plasma frequency and penetration depth. Furthermore, the gravitomagnetic-like tensor field is predicted to be expelled from the superconductor in a gravitational Meissner-like effect and a charge separation effect in the superconductor is to be expected as a result of the gravitational wave.

Going beyond the linear approximation to the Einstein field equations, we are in a position to evaluate the existence of gravitational wave reflection from the standpoint of gravitational theory itself. 

Let us model a static gravitating source as a perfect fluid with vanishing pressure, that is particulate-like. We can approximate a lot of matter sources with this model including a solid bulk metal with density $\rho$. In the rest frame of the material the stress-energy tensor is diagonal with only one non-vanishing component $T_{\sigma \nu}={\rm diag} (\rho, 0, 0, 0)$. It is exactly this matter configuration which produces the Newtonian limit to the Einstein field equations, therefore the metric for the static Newtonian sources is given by
\begin{eqnarray}\label{Newtonian}
	ds^2=-\left(1+ \frac{2\Phi}{c^2} \right) d x_0^2 + \left(1- \frac{2\Phi}{c^2} \right) (dx^2 + dy^2 +dz^2),   
\end{eqnarray}    
where $x_0 = ct$ and $\Phi(\vec{r})=-GM/r$. Here $G$ is the gravitational constant, $M$ is the mass of the body that creates the gravitational potential and $r=\sqrt{x^2+y^2+z^2}$ is radial distance to the source. An expression for $\Phi$ in the case of distribution of masses with density $\rho$ can also be given. 

Now we insert the metric (\ref{Newtonian}) into the curved space-time d'Alambertian (\ref{GW_eqn}) to obtain 
\begin{eqnarray}\label{eqn_Newton}
	&&\frac{g^{00}}{g^{xx}} \; \partial_0^2 \phi + \partial_x^2	\phi + \partial_y^2	\phi + \partial_z^2	\phi +   \frac{\mu^{2}}{g^{xx}} \; \phi \\
	\nonumber &&\qquad + \left\{ \partial_x  A \right\} \partial_x  \phi + \left\{ \partial_y  A \right\} \partial_y  \phi + \left\{ \partial_z  A \right\} \partial_z  \phi =0,
\end{eqnarray}    
where the effective mass is given by $\mu^{2} = \left\{ \frac{\kappa}{2} \rho c^2 +  \Lambda \right\}$, $g^{00}= - (1+\epsilon)^{-1}$, $g^{xx}=  (1-\epsilon)^{-1}$, $\sqrt{-g}=(1-\epsilon)\sqrt{1-\epsilon^2}$ and $|\epsilon|=2|\Phi|/c^2 \ll 1$ is a small parameter. Here $A=\ln{g^{xx}} + \ln{\sqrt{-g}}=\ln{\sqrt{1-\epsilon^2}}$. Note the particular simplification in equation (\ref{eqn_Newton}) is possible by virtue of $g^{xx}=g^{yy}=g^{zz}$.

Next we insert the ansatz 
\begin{eqnarray}\label{ansatz}
	\phi=e^{-\frac12 \int \nabla A \cdot d\vec{r} } \;  \chi
\end{eqnarray}    
to remove the terms proportional to first derivatives at the expense of introducing an effective geometric potential $V_{g}=-(\nabla A)^2/4$,
\begin{eqnarray}\label{GW_eqn_Newton}
	\left[ \frac{g^{00}}{g^{xx}} \; \partial_0^2  + \partial_x^2	 + \partial_y^2	 + \partial_z^2 \right] \chi	 +   \left[ -\frac14 (\nabla A)^2 + \frac{\mu^{2}}{g^{xx}} \right] \; \chi = 0.
\end{eqnarray}

The term multiplying the temporal second order derivative has the meaning of an index of refraction $n=c/v$:
\begin{eqnarray}\label{index_refraction}
	\frac{g^{00}}{g^{xx}} = - \frac{1-\epsilon}{1+\epsilon}=-n^2,
\end{eqnarray}    
therefore the velocity of propagation of these gravitational waves is given by
\begin{eqnarray}\label{velocity_propagation}
	n = \sqrt{\frac{1-\epsilon}{1+\epsilon} } \Rightarrow v=c\sqrt{\frac{1+\epsilon}{1-\epsilon} } < c.
\end{eqnarray}    
Since $\epsilon < 0$, the velocity of propagation is less than the velocity of light in vacuum.

The effective geometric potential comes out as a second order correction in terms of the small parameter
\begin{eqnarray}\label{V_geom}
	V_{g}=-\frac14 (\nabla A)^2 = - \frac{1}{16} \frac{\epsilon^2}{(1-\epsilon^2)^2} (\nabla \epsilon)^2.
\end{eqnarray}    
Expanding along the small parameter and substituting $\epsilon$ with the expression containing the gravitational potential
\begin{eqnarray}\label{}
	V_{g}\approx -\frac{1}{c^8} \Phi^2 \left(\nabla \Phi\right)^2 = - \frac{G^4 M^4 }{c^8 r^6},
\end{eqnarray}    
which is negligibly small. 

However, in the case which interests us, that is the boundary between gravitationally dense and sparse (including vacuum) materials the density gradient matters $\nabla \epsilon \gg 1 $
and we can model the geometric potential as being proportional to the Dirac delta function if the density is a step-wise function, that is there is a {\it sharp boundary} between materials with large density difference. Suppose, we have a very dense slab with thickness $d$ and cross-section $S$, then we may put
\begin{eqnarray}\label{V_geom_equiv}
	V_{g}\approx  - \frac{G^4 \rho^4 S^4}{c^8 d^2} \delta(\vec{r}).
\end{eqnarray}

The mass term is a constant in the first order approximation $\frac{\mu^{2}}{g^{xx}}= \mu^{2} -\epsilon \; \mu^{2}$
and negligible due to being proportional to the Einstein constant. Interestingly, it is dominating over the geometric potential (\ref{V_geom_equiv}). 

Finally, we can set the toy model for exploring the reflection and transmission properties of gravitational waves
\begin{eqnarray}\label{GW_toy_model}
	\left[ -n^2 \; \partial_{ct}^2  + \Delta + \mu^2 \right] \; \chi (t,\vec{r}) = 0.
\end{eqnarray}
Here $\Delta$ is the flat space Laplacian, which can be cast into spherical coordinates as well. The index of refraction $n^2 \approx 1 - 2 \epsilon$ (\ref{index_refraction}) and the effective mass $\mu$  can be assumed constants. 

The index of refraction $n$ can therefore be set to
\begin{eqnarray}\label{index_of_refraction}
	n = 1-\frac{2\Phi}{c^2} 
\end{eqnarray}
and depends on the local Newtonian gravitational potential $\Phi$:
\begin{eqnarray}\label{newton_potential}
	\Phi(\vec{r})=-\int_{\mathbb{R}^3} G \frac{ \rho(\vec{r'}) }{|\vec{r}-\vec{r'}|} dV',
\end{eqnarray}
where $G$ is Newton's gravitational constant, $\rho$ is the local material density. Obviously, the solution for a constant density mass distribution is $\Phi=-G\rho V/r$, where $V$ is the Euclidean volume the material takes up.

Now suppose we have alternating layers with density $\rho_1$ and $\rho_2$, then the difference in the indices of refraction is given by the difference in their densities:
\begin{eqnarray}\label{}
	n_2-n_1 = 2(\Phi_1-\Phi_2)/c^2 = 2V(\rho_2 - \rho_1)/rc^2 . 
\end{eqnarray}
Alternatively, $n_2+n_1 = 2$. Here, it is obvious that in order to increase the difference in the index of refraction between materials, we have to aim at combining materials with vastly different densities. Examples will be discussed in Section \ref{sec:propulsion}.

\subsection{Boundary Conditions}

The Fresnel equations in optics describe the reflection and transmission of electromagnetic waves incident on an interface between materials with different indexes of refraction. They are a result of the boundary conditions for the electric and magnetic fields at the interface, which in their own right are a consequence of the Maxwell's equations. In the case of the electromagnetic waves, the magnetic field is proportional to the electric field and one can assume as fundamental only the two boundary conditions for the tangential $E_{t}$ and normal $E_{n}$ to the interface electric fields
$E^{(1)}_{t}=E^{(2)}_{t}$ and $\varepsilon_1 E^{(1)}_{n} = \varepsilon_2 E^{(2)}_{n}$.

Einstein field equations in the weak field limit can be cast into a form identical to Maxwell's equations. We are going to explore this analogy by translating the optical Fresnel reflection coefficient for normal incidence in the case of gravitational waves. As a result, we may expect a relation of the type
\begin{eqnarray}\label{}
	R=\left| \frac{n_1-n_2}{n_1+n_2} \right| = V|\rho_1 - \rho_2|/rc^2,   
\end{eqnarray}
which is very small but non-zero, $0<R \ll 1 $. Here we will assume that if the order of the two media is changed, then $R_{21}=R_{12}=R$ as dictated by the above formula. The transmission coefficient can be defined as
$T_{12}=1-R_{12}=2n_2/(n_1+n_2)\approx 1$ and $T_{21}=1-R_{21}=2n_1/(n_1+n_2)\approx 1$. Therefore $T=T_{12}=T_{21}=1$.

When we have such a small value for the reflection coefficient, we may add additional (say $n$) alternating interfaces with the effect of having these reflection coefficients
\begin{eqnarray}\label{}
	\nonumber	R_1&=& R, \\
	\nonumber	R_2&=& R + T^2 R \approx 2 R, \\
	\nonumber	R_3&=& R + T^2 R + T^4 R \approx 3 R,\\
	\nonumber	R_4&=& R + T^2 R + T^4 R + T^6 R \approx 4 R, \\
	\nonumber & \cdots& \\
	R_n & \approx& n R,
\end{eqnarray}
add up in a linear progression. If we want to increase the reflection coefficient for gravitational waves hundred times, we have to stack hundred interfaces between materials with vastly different densities.

Let us try a second approach.  We will take the gravitational wave equation (\ref{GW_eqn}) in its weak field limit (\ref{GW_toy_model}) as a stand alone partial differential equation and assume that the appropriate boundary conditions come as a result of the gluing procedure for the solutions (and their derivatives) on either side of the interface. Such boundary conditions are typical in quantum mechanics and therefore we may expect quantum mechanics-like behavior for the reflection and transmission coefficients.

Let us consider a traveling plane wave solution to (\ref{GW_toy_model}), that is $\chi=A \exp{i(\omega t \pm \vec{k} . \vec{r})}$, where $A$ is a constant, $\omega$ is the circular frequency and $\vec{k}$ is the wave vector. It converts the differential equation into an algebraic one with a solution
\begin{eqnarray}\label{}
	\frac{n^2 \omega^2}{c^2} + \mu^2 = k^2,
\end{eqnarray} 
turning the plane wave into a genuine solution to the differential equation (\ref{GW_toy_model}). Now, suppose the reflecting interface is at $x=0$ ($y=z=0$ the coordinate system is centered on the interface) and we consider a plane wave impinging on the left hand side (subscript 1), gets partially reflected back and penetrates on the right hand side (subscript 2). In this case the solution on the left hand side has the form
	\begin{eqnarray}\label{}
		\chi_1 (t,\vec{r}) = e^{i(\omega t - \vec{k}_1 . \vec{r})} + R \; e^{i(\omega t + \vec{k}_2 . \vec{r})} .
	\end{eqnarray}
	On the right hand side  of the interface the solution is given by 
	\begin{eqnarray}\label{}
		\chi_2 (t,\vec{r}) = T \; e^{i(\omega t - \vec{k}_2 . \vec{r})} .
	\end{eqnarray}

Note, the wavenumbers change due to the different indices of refraction and mass on either side of the interface. The $x$-derivative of these solutions, provided the propagation is only along the $x$ direction, that is $\vec{k}=(k,0,0)$, are given by
	\begin{eqnarray}\label{}
		\partial_x \chi_1 (t,\vec{r}) = - i k_1 \left[ e^{i(\omega t - \vec{k}_1 . \vec{r})} - R \; e^{i(\omega t + \vec{k}_1 . \vec{r})} \right],
	\end{eqnarray}
	where $k_1^2=\mu_1^2+n_1^2 \omega^2 /c^2$ and
	\begin{eqnarray}\label{}
		\partial_x \chi_2 (t,\vec{r}) = - i k_2 T  e^{i(\omega t - \vec{k}_2 . \vec{r})},
	\end{eqnarray}
where $k_2^2=\mu_2^2+n_2^2 \omega^2 /c^2$. We now impose the standard gluing conditions for the solutions and their derivatives on the interface
	\begin{eqnarray}\label{}
		\chi_1 (0,0) &=& \chi_2 (0,0),\\
		\partial_x \chi_1 (0,0) &=& \partial_x \chi_2 (0,0),
	\end{eqnarray}
to obtain the system of equations for the reflection coefficient $R$ when a monochromatic gravitational wave traverses the interface from the first medium into the second medium
	\begin{eqnarray}\label{}
		1+R &=& T,\\
		k_1 (1-R) &=& k_2 T = k_2 (1+R),
\end{eqnarray}
where the first relation is the law of conservation of energy and upon inserting it into the second relation we get
\begin{eqnarray}\label{}
	R_{12} &=& \frac{k_1-k_2}{k_1+k_2} = \frac{n_1-n_2}{n_1+n_2},
\end{eqnarray}
where we have assumed that $\mu_1 \approx 0$ and $\mu_2 \approx 0$ in order to expand $k_1$ and $k_2$ and have inserted a subscript 12 to denote the order of media in the gravitational wave's path. 

Let us check if the derivation is symmetric when it comes to the choice of which side of the interface the gravitational wave impinges upon. If we started with side 2, insted of 1, we would have to change the indices of the solutions and the signs of the wavevectors $\vec{k}_1 \to - \vec{k}_1$ and $\vec{k}_2 \to - \vec{k}_2$ to obtain the system 
	\begin{eqnarray}\label{}
		1+R &=& T,\\
		k_2 (1-R) &=& k_1 T = k_1 (1+R),
\end{eqnarray}
with the solution
\begin{eqnarray}\label{}
	R_{21} &=& \frac{k_2-k_1}{k_1+k_2} = \frac{n_2-n_1}{n_1+n_2}.
\end{eqnarray}
Therefore, the solution is anti-symmetric when it comes to the change in the order of media in gravitational wave's path
\begin{eqnarray}\label{}
	R_{12} &=& -R_{21}.
\end{eqnarray}
In this case, stacking multiple layers of materials with alternating density is not going to lead to an increase in the reflection coefficient:
\begin{eqnarray}\label{}
	\nonumber	R_1&=& R_{12}, \\
	\nonumber	R_2&=& R_{12} + T^2 R_{21} \approx 0, \\
	\nonumber	R_3&=& R_{12} + T^2 R_{21} + T^4 R_{12} \approx R_{12},\\
	\nonumber	R_4&=& R_{12} + T^2 R_{21} + T^4 R_{12} + T^6 R_{21} \approx 0, \\
	\nonumber & \cdots& \\
	R_n & \approx& \left\{ R_{12} \; {\rm for} \; n=2l+1 \quad|\quad 0 \; {\rm for} \; n=2l \right\},
\end{eqnarray}
where we have assumed that $T_{12}=T_{21}=T\approx 1$. This assumption is valid for small $|R| \ll 1$. Note the cancellation of the reflection coefficient when an even number of interfaces, that is an odd number of layers is involved. This poses a very clear experimental outcome which can settle the issue of what are the proper boundary conditions involved in gravitational wave reflection.

The difference with the Fresnel-like case stems from the choice of the boundary conditions or the inappropriate understanding of the properties of gravitational wave reflection. In the case when the gravitational wave undergoes a $\pi$-phase shift when reflecting off a denser substrate (say $n_1 > n_2$, no phase shift takes place when going from 1 to 2), then going from 2 to 1 will introduce a minus sign in front of the reflection coefficient $R \to -R$, effectively setting 
\begin{eqnarray}\label{optical_b.c.}
	R_{12}=R_{21} &=& \left| \frac{n_1-n_2}{n_1+n_2} \right|.
\end{eqnarray}

The correct interpretation of the properties of gravitational wave reflection thus remains an open question which may find a resolution in an experimental or observational setting. Such an experimental setting can arise while looking at the two cases. Provided one stacks an even number of interfaces and tries to reflect gravitational waves off of such a mirror, the resolution to the issue would come out automatically, since the two cases are vastly different. One prediction gives reflection, while the other prediction indicates no reflection (even number of interfaces is equivalent to odd number of layers). It is a binary experiment, which is very easy to interpret. 

In what follows, we will assume the ``optical-like'' boundary conditions (\ref{optical_b.c.}) hold for gravitational wave reflection and one can construct a sufficiently effective mirror by stacking layers of materials with vastly different densities. The assumption does not validate or invalidate the arguments to follow, but merely increases the odds of a successful experimental implementation.

\section{Propulsion}\label{sec:propulsion}

Suppose we have a source of gravitational waves, that is a gravitational wave emitter (GWE). This source may be controllable in the case of some other form of energy, say electrical, being converted into gravitational wave emission or may be of natural origin, say a spiraling inwards neutron star binary.

The GWE should be stationary or in constant uniform motion (in an inertial frame) during the emission process. Regardless of the emission diagram, it should be symmetric and not create an uncompensated momentum in any direction. For all we experimentally know, there is no instance of violation of Newton's third law of Mechanics, that is: to every action one should find an equal and opposite reaction. The law is explicitly stated in terms of balance of forces. In effect, there is no motion of the center of mass of the GWE during the emission process. 

Indeed, in the case of gravitation, which according to Einstein field equations is not a force per se, but rather a curvature in space-time fabric, it is not so clear whether Newton's third law of Mechanics holds. We argue that it holds in the case of  Einsteinean gravitation as well. Since Newton's third law of Mechanics is the necessary and sufficient condition for the law of conservation of energy to hold \cite{Newton}, we argue that since Einstein field equations follow from an action principle (interpretation of the law of conservation of energy) \cite{Hilbert, Feynman}, they inherit ``the balance of forces'' encoded in Newton's third law of Mechanics. As a result, gravitation in all its formulations is compatible with Newton's third law of Mechanics by virtue of its derivation. Therefore, we cannot ever expect to get propulsion using the gravitational field alone, just like we do not produce propulsion using a chemical explosion alone.   

The classical mechanics case is very clear. Think of a rocket engine. One has to ``reflect'' the momentum of the particles in the hot high speed propulsive jet off of the exhaust chamber/nozzle and in effect eject mass rearward, to produce a thrust forward. 

In the case of gravitation, producing ``thrust'' follows the same principle, only this time around the gravitational wave traveling in a given direction should be partially reflected off of a gravitational wave mirror which serves as an ``exhaust'' nozzle breaking the symmetry of the GWE emission diagram. Thus an excess of gravitational wave density (curvature) is created at one side and the GWE follows the curvature by constantly falling, that is following a geodesic world line with spatial component. Without the gravitational wave mirror, the world line is time-like, that is the GWE stays stationary in space and propagates its history only in time. 

As discussed in the previous section, the gravitational wave mirror looks like a stacked layered structure made up of a collection of  {\it sharp boundaries} between materials with large density difference. Possible combinations include Material I (magnesium 1.74 g/cm$^3$, beryllium 1.85 g/cm$^3$, aluminum 2.71 g/cm$^3$ or an organic epoxy based layer $\approx$ 1 g/cm$^3$) Material II (uranium 18.95 g/cm$^3$, gold 19.32 g/cm$^3$, tungsten 19.35 g/cm$^3$, platinum 21.45 g/cm$^3$, iridium 22.4 g/cm$^3$). The optimal combination such as Mg-Ir may not be possible due to adhesion issues stemming from the mismatch in the crystal structure, hexagonal close-packed (hcp) in the case of magnesium versus face-centered cubic (fcc) in the case of iridium \cite{kittel}. However, the combination Al-Ir is quite favorable due to both metals having a fcc crystal lattice with matching cell parameters $a_{Mg}=405$ pm and $a_{Ir}=384$ pm \cite{ir}. 

\begin{figure}[t]
	\begin{center}
		\includegraphics[scale=2]{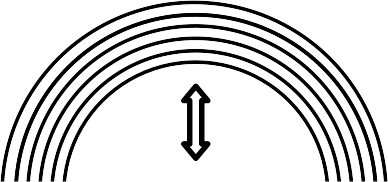}
		\caption{\label{fig:mirror} The propulsion scheme proposed here consists of a GWE, i.e. a quadrupole mass oscillating device hence the opposing arrow directions, placed in the center of a hemisphere shaped gravitational wave deflector made of layered materials with large density differential such as aluminum and iridium. 
		}
	\end{center}
\end{figure}

It is now not hard to imagine a gravitational wave propulsion device which should consist of an emitter and a deflector (see Figure \ref{fig:mirror}). The deflector's shape is a hemisphere with the intention to break the symmetry of the emission diagram. The emitter is an oscillating quadrupole mass distribution which can be realized in the solid state as well by placing two superconducting tunnel junctions on a line and driving them from a contact in between them. The electrical driving scheme is discussed elsewhere \cite{AS, VA}.

\section{Conclusions}

In conclusion, let us summarize the main results of the paper: (i) the equation for the amplitude of the gravitational waves beyond the linear approximation is the massive scalar Klein-Gordon equation in which the mass of the field is given by a combination of the mass-energy density of matter and the cosmological $\Lambda$ term; (ii) the equation in the case of a Schwarzschild black hole can be cast into an effective Schr\"odinger form where the effective geometric potential is negative (bound states) for the zero angular momentum state and repulsive for the non-zero angular momentum states outside the event horizon pointing to gravitational wave reflection taking place. This is an unambiguous experimental prediction of black hole existence -- scattering of gravitational waves from the event horizon; (iii) the reduction of the gravitational wave equation to the Newtonian limit leads to the emergence of an index of refraction, therefore the possibility, by virtue of an analogy with the Fresnel equations or quantum mechanical gluing procedure, of reflecting some of the incident gravitational wave power; (iv) gravitational wave mirror according to this result consists of a layered structure of interfaces with sharp density boundary, thus opening up gravitational wave propulsion physics; and finally (v) we discuss this type of propulsion in the light of Newton's third law of Mechanics. 

\section*{Acknowledgements}

The authors wrote the paper without the use of LLMs or other types of AI. No new data has been created. Author contributions are as follows: V. A. devised and wrote the initial manuscript with valuable inputs, discussion, comments and edits from A. S. The work of A. S. at Los Alamos National Laboratory was carried out under the auspices of the U.S. DOE and NNSA under Contract No. DEAC52-06NA25396. Fruitful discussions with Ralf Sch\"utzhold are also acknowledged.

\end{document}